\documentclass[a4paper]{article}

\usepackage[utf8]{inputenc}

\usepackage{multicol}
\usepackage{amsthm}
\usepackage{bm}
\usepackage{physics}

\usepackage[dvipdfmx]{graphicx}

\usepackage{bmpsize}
\usepackage{amssymb}
\usepackage{amsmath}
\usepackage{color}
\usepackage{amsthm}
\usepackage{comment}

\newcommand{\BA}{\begin{eqnarray}}
\newcommand{\EA}{\end{eqnarray}}

\definecolor{dgreen}{rgb}{0.0, 0.5, 0.0}

\begin{document}

\fontsize{14pt}{16.5pt}\selectfont

\begin{center}
\bf{
Ultradiscretization in discrete limit cycles \\
of tropically discretized and max-plus Sel'kov models
}
\end{center}
\fontsize{12pt}{11pt}\selectfont
\begin{center}
Yoshihiro Yamazaki$^{1 *)}$ and Shousuke Ohmori$^{2, 3}$\\ 
\end{center}

\vspace{1mm}

\noindent
$^1$\it{Department of Physics, Waseda University, Shinjuku, Tokyo 169-8555, Japan}\\
$^2$\it{National Institute of Technology, Gunma College, Maebashi-shi, Gunma 371-8530, Japan}\\
$^3$\it{Waseda Research Institute for Science and Engineering, Waseda University}{Shinjuku, Tokyo 169-8555, Japan}\\

\noindent
*corresponding author : yoshy@waseda.jp\\
~~\\
\rm
\fontsize{11pt}{14pt}\selectfont\noindent

\baselineskip 30pt

\noindent
{\bf Abstract}\\
%\begin{abstract}
%
%
The state of limit cycles for a tropically discretized Sel'kov model 
becomes ultradiscrete due to phase lock caused by 
a saddle-node bifurcation. 
This property is essentially the same 
as the case of the negative feedback model, 
and existence of a general mechanism for 
ultradiscretization of the limit cycles is suggested.
Furthermore in the case of the max-plus Selkov model, 
we find the logarithmic dependence 
of the time to pass the bottleneck 
for phase drift motion in the vicinity 
of the bifurcation point.
This dependency can be understood 
as a consequence of the piecewise linearization 
by applying the ultradiscrete limit.

\bigskip

\bigskip

\bigskip

%\end{abstract}

%\noindent
%{\bf Key Words} : {ultradiscretization, limit cycle, 
%phase lock, saddle-node bifurcation, bottleneck motion}\\

%\section{Introduction}
%\label{sec:1}

%
%%%%%%%%%%%%%%%%%%%%%%%%%%%%%%%%%%%%%%%%%%%%%%%%%%%%%%%%%%%%%%%%%%%%%%%%%%%%%%%%%%%%%%%%%%%%%%%%%%%%%%%%%%
%
%%%%%%%%%%%%%%%%%%%%%%%%%%%%%%%%%%%%%%%%%%%%%%%%%%%%%%%%%%%%%%%%%%%%%%%%%%%

\baselineskip 24pt

Recently, we have reported the results of numerical analysis 
for the dynamical properties of the tropically discretized 
negative feedback model \cite{Yamazaki2023,Ohmori2023c,Gibo2015}.
This model includes a positive parameter $\tau$, 
which corresponds to the time interval for discretization.
Discrete stable (attractive) limit cycle solutions 
emerge in this model when $\tau$ is larger 
than a finite positive value $\tau_{0}$. 
The interesting point is that the states (phases) 
of the limit cycles become ultradiscrete 
due to phase lock by saddle-node bifurcation 
at $\tau = \tau^{\ast} > \tau_{0}$.
Additonally, there exist unstable (repulsive) limit cycles in addition to stable ones 
when $\tau > \tau^{\ast}$.
Furthermore the existence of these stable and unstable 
ultradiscrete limit cycles is retained 
even in the max-plus model, 
which are obtained from the discretized model 
in the ultradiscrete limit\cite{Tokihiro2004}.

\begin{figure}[t]%[h!]
	\begin{center}
	% \includegraphics[bb=0 0 960 720, width=6.5cm]{Fig1a.png}
	% \hspace{-20mm}
	\includegraphics[width=7.5cm]{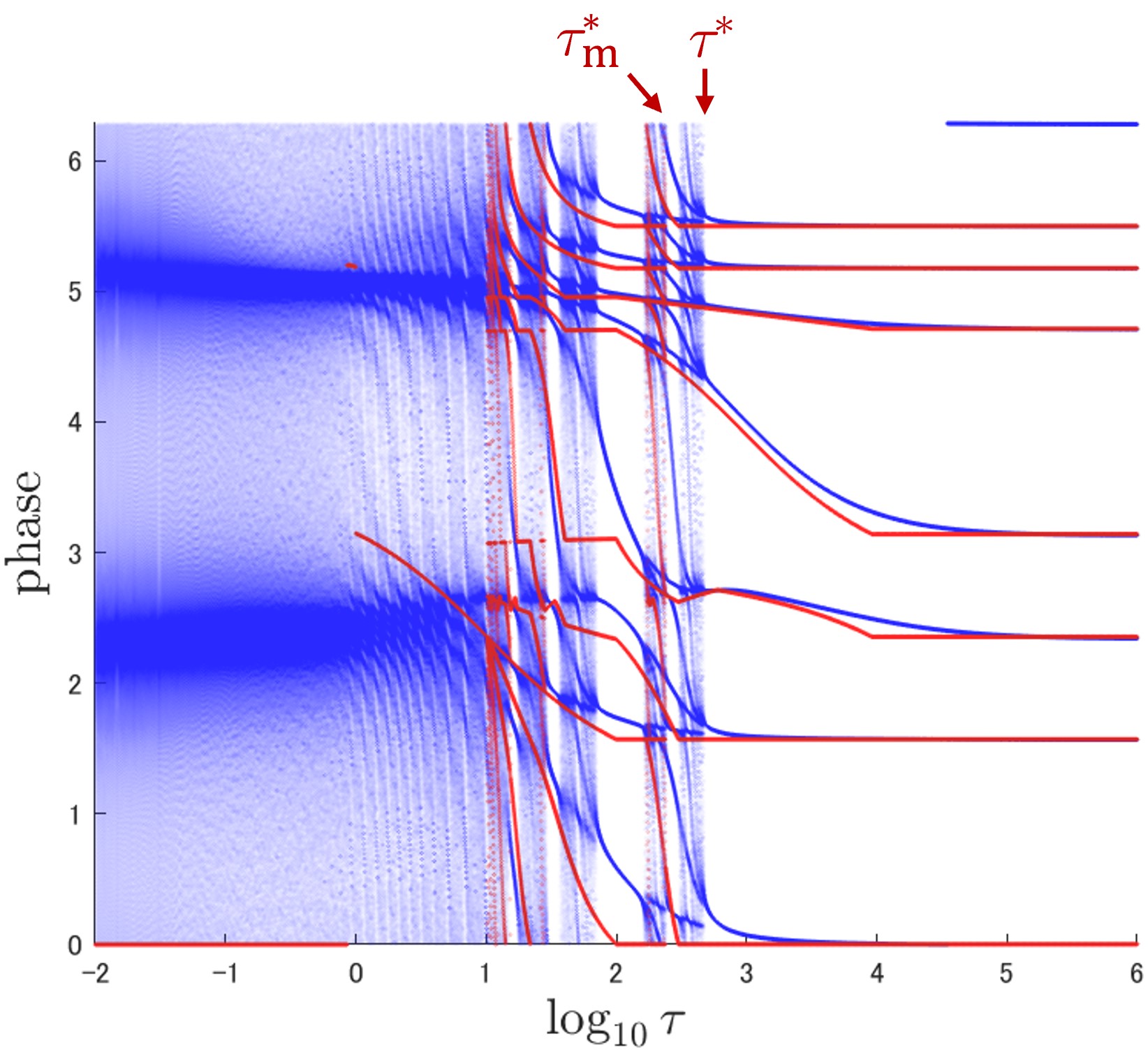}
	\caption{\label{fig:BifDiagram_Selkov_ul_ov} 
		The blue scatter plot: the bifurcation diagram 
		for the phase $\theta_{n}$ as a function of $\tau$ 
		obtained from eq.(\ref{eqn:Selkov_tropical}) 
		with $a=0.01$ and $b=0.98$.
		The red: the max-plus bifurcation diagram 
		for $\Theta_{n}(\tau)$ obtained from eq.(\ref{eqn:Selkov_maxplus}).}
	\end{center}
\end{figure}

So far the above dynamical properties have been confirmed 
only in the negative feedback model. 
It is unclear whether they hold only 
for the negative feedback model or more generally.
In this letter, with a focus on their generality, 
we demonstrate another example, 
the tropically discretized Sel'kov model\cite{Ohmori2021,Yamazaki2021,Ohmori2022}, 
\begin{equation}
	\left\{
		\begin{aligned}
			x_{n+1} & = \frac{x_n+\tau(ay_n+x^2_ny_n)}{1+\tau} \equiv \eta(x_n, y_n), \\
			y_{n+1} & = \frac{y_n+\tau b}{1+\tau(a+x_n^2)} \equiv \xi(x_n, y_n). 
		\end{aligned}
	\right.
	\label{eqn:Selkov_tropical}
\end{equation}
Equation (\ref{eqn:Selkov_tropical}) can be derived 
from the following continuous model\cite{Selkov1968,Strogatz15} 
via the tropical discretization\cite{Murata2013,Matsuya2015}
with the additional positive parameter $\tau$ 
for the discrete time step, 
\begin{equation}
	\left\{
		\begin{aligned}
			\frac{dx}{dt} & = -x+ay+x^2y, \\
			\frac{dy}{dt} & = b-ay-x^2y, 
		\end{aligned}
	\right.
	\label{eqn:Selkov_continuous}
\end{equation}
where $x$, $y$, $a$, and $b$ are positive.
We have already reported 
that eq.(\ref{eqn:Selkov_tropical}) has 
limit cycle solutions for all $\tau$ 
when we set $a = 0.01$ and $b = 0.98$
\cite{Ohmori2021,Ohmori2023b}.
Defining the phase $\theta_{n}(\tau) \in [0, 2\pi)$ 
in the limit cycles as 
\begin{equation}
	\theta_{n}(\tau) = \arctan 
	\displaystyle \frac{\ln y_{n}-\ln \bar{y}}{\ln x_{n}-\ln \bar{x}}, 
	\label{eqn:def_theta}
\end{equation}
where $(\bar{x}, \bar{y})$ is the fixed point of eq.(\ref{eqn:Selkov_tropical}), 
we obtain the bifurcation diagram of $\{ \theta_{n}(\tau) \}$  
shown as the blue scatter plot in Fig.\ref{fig:BifDiagram_Selkov_ul_ov}.
It is clearly found that distribution of $\{ \theta_{n}(\tau) \}$ 
changes at $\tau = \tau^{\ast}$ 
and that the state of the limit cycle becomes ultradiscrete 
when $\tau > \tau^{\ast}$.

\begin{figure}[t]%[h!]
	\begin{center}
	% \includegraphics[bb=0 0 960 720, width=6.5cm]{Fig1a.png}
	% \hspace{-20mm}
	\includegraphics[width=7.5cm]{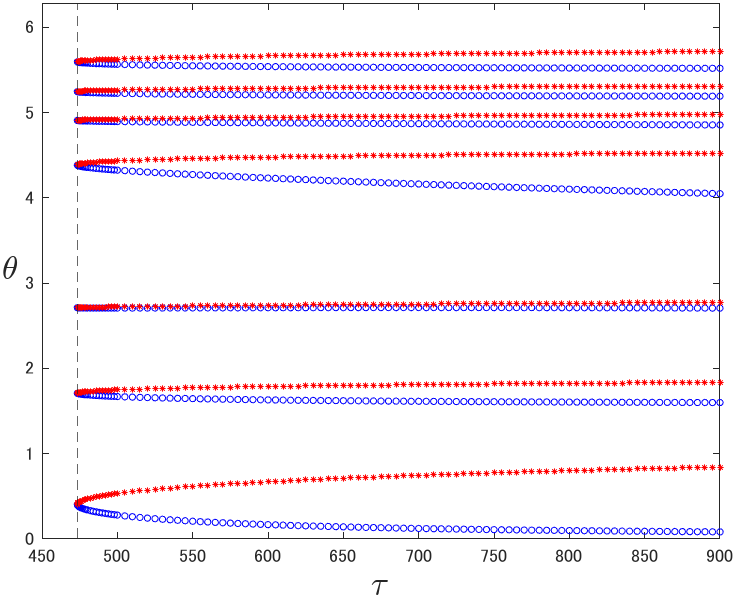}
	\caption{\label{fig:Phase_st_unst_all_Selkov}
		The phases $\bar{\theta}_{7}$ for the fixed points 
		$(\bar{x}_{7}, \bar{y}_{7})$ of eq.(\ref{eqn:Selkov_tropical_7}) 
		as a function of $\tau$.}
	\end{center}
\end{figure}

The blue plot in Fig.\ref{fig:BifDiagram_Selkov_ul_ov} 
shows that the ultradiscrete limit cycle 
for $\tau > \tau^{\ast}$ consists of seven states.
(For more discussion regarding the number of states, see ref.\cite{Yamazaki2021}.)
Here we consider the time evolution of the state at every seven steps, 
\begin{equation}
	x_{7(n+1)}  =  \eta^{7}(x_{7n}, y_{7n}), \qquad
	y_{7(n+1)}  =  \xi^{7}(x_{7n}, y_{7n}), 
	\label{eqn:Selkov_tropical_7}
\end{equation}
where $\eta^{7}$ and $\xi^{7}$ show 7-th iterates of 
$\eta$ and $\xi$ in eq.(\ref{eqn:Selkov_tropical}).  
We introduce the phase $\bar{\theta}_{7}$
of the fixed point $(\bar{x}_{7}, \bar{y}_{7})$ 
from eq.(\ref{eqn:def_theta}), 
where $(\bar{x}_{7}, \bar{y}_{7})$ satisfies  
$\bar{x}_{7}  =  \eta^{7}(\bar{x}_{7}, \bar{y}_{7})$ and  
$\bar{y}_{7} = \xi^{7}(\bar{x}_{7}, \bar{y}_{7})$.
Based on eq.(\ref{eqn:Selkov_tropical_7}), 
we can numerically estimate the values of $\tau^{\ast}$ 
as $\tau^{\ast} \approx 473.439297\cdots$ 
by the upper limit value for absence of the fixed points.
We obtain 14 fixed points when $\tau > \tau^{\ast}$ 
and the phases obtained from them are shown 
in Fig.\ref{fig:Phase_st_unst_all_Selkov} 
as a function of $\tau$. 
It is noted that 7 phases with the blue open circles, 
denoted by $\left\{\bar{\theta}_{7}^{\rm{(s)}}\right\}$ hereafter, 
are identical to the ultradiscrete states 
for $\tau > \tau^{\ast}$ shown 
by the blue plot in Fig.\ref{fig:BifDiagram_Selkov_ul_ov}.
The other 7 phases with the red asterisks 
are denoted by $\left\{\bar{\theta}_{7}^{\rm{(u)}}\right\}$.

\begin{figure}
	\centering
	\begin{minipage}{0.45\textwidth}%{0.23\textwidth}
			\includegraphics[width=\textwidth]{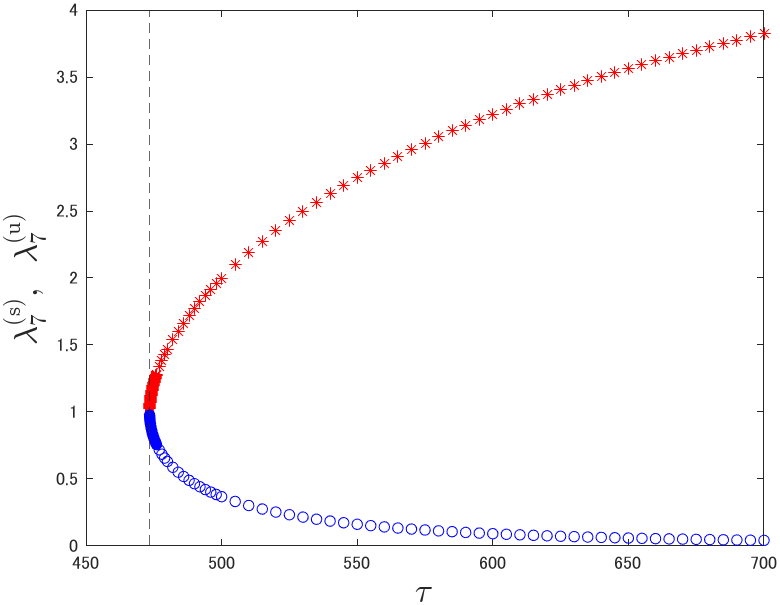}
			\centering
			(a)
	\end{minipage}
	%\hfill
	\begin{minipage}{0.45\textwidth}%{0.23\textwidth}
			\includegraphics[width=\textwidth]{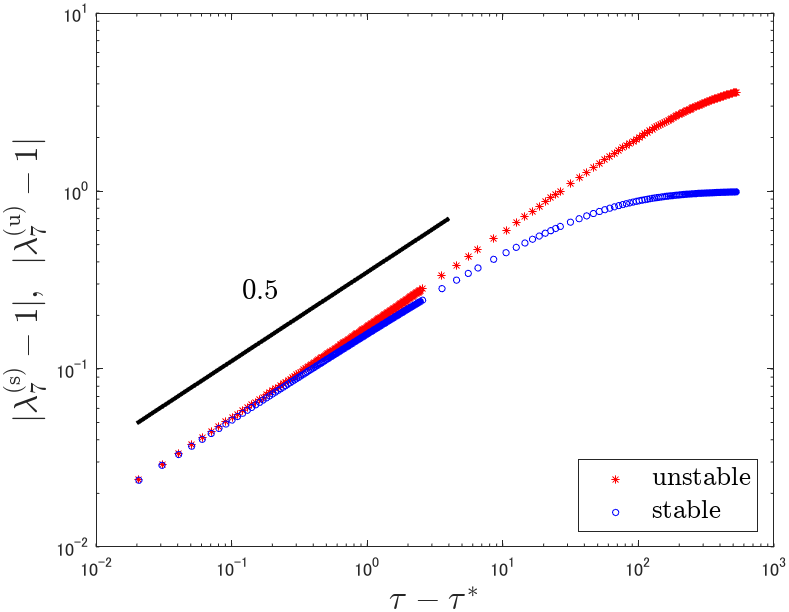}
			\centering
			(b)
	\end{minipage}
	\caption{\label{fig:Lambda_st_unst_lin_Selkov} 
		(a) The maximum eigenvalues 
		$\lambda_{7}^{\text{(s)}}$ (blue circle) 
		and $\lambda_{7}^{\text{(u)}}$ (red asterisk) 
		of the Jacobi matrix for the fixed points 
		$\{\bar{\theta}_{7}^{\rm{(s)}}\}$ 
		and $\{\bar{\theta}_{7}^{\rm{(u)}}\}$, respectively.
		(b) The scaling relations between 
		$\lambda_{7}^{\text{(s)}}$, $\lambda_{7}^{\text{(u)}}$ 
		and $\tau - \tau^{\ast}$.}
\end{figure}

For the stability of these fixed points, 
the eigenvalues of their Jacobi matrix are focused on.
Figure \ref{fig:Lambda_st_unst_lin_Selkov}(a) shows 
the maximum eigenvalues $\lambda_{7}^{\rm(s)}$ 
(blue circles) and $\lambda_{7}^{\rm(u)}$ 
(red asterisks) for $\left\{\bar{\theta}_{7}^{\rm{(s)}}\right\}$ 
and $\left\{\bar{\theta}_{7}^{\rm{(u)}}\right\}$, respectively.
It is concluded from this figure that 
$\left\{\bar{\theta}_{7}^{\rm{(s)}}\right\}$ are stable 
and $\left\{\bar{\theta}_{7}^{\rm{(u)}}\right\}$ are unstable, 
and that saddle-node bifurcation occurs at $\tau=\tau^{\ast}$.
In addition, Fig.\ref{fig:Lambda_st_unst_lin_Selkov}(b) 
shows the asymptotic property for $\tau$ dependence 
of $\lambda_{7}^{\rm{(s)}}$ and $\lambda_{7}^{\rm{(u)}}$: 
$|\lambda_{7}^{\rm{(s)}}-1|, |\lambda_{7}^{\rm{(u)}}-1| \sim 
(\tau - \tau^{\ast})^{0.5}$.
These dynamical properties for ultradiscretization 
of the Sel'kov model are essentially the same 
as the case of the negative feedback model, 
though the number of the ultradiscrete states is different.
%

%%%%%%%%

Applying the variable transformations, 
$\tau = e^{T/\varepsilon}$, $x_n = e^{X_n/\varepsilon}$, 
$y_{n}= e^{Y_{n}/\varepsilon}$, $a  = e^{A/\varepsilon}$, 
$b =e^{B/\varepsilon}$, 
and then the ultradiscrete limit\cite{Tokihiro2004}, 
$$ 
  \lim_{\varepsilon \to 0} \varepsilon \ln 
  \left( e^{\frac{P}{\varepsilon}} + e^{\frac{Q}{\varepsilon}}
  + \cdots \right) = \max (P, Q, \cdots), 
$$
to eq.(\ref{eqn:Selkov_tropical}), we obtain 
\begin{equation}
	\left\{
		\begin{aligned}
			X_{n+1} & = \max(X_n, T+\max(A+Y_n,2X_n+Y_n))-\max (0,T), \\
			Y_{n+1} & = \max(Y_n,T+B)-\max(0,T+\max(A,2X_n)).
		\end{aligned}
	\right.
	\label{eqn:Selkov_maxplus}
\end{equation}
Between the variables in eq.(\ref{eqn:Selkov_tropical}) and eq.(\ref{eqn:Selkov_maxplus}), 
the following relations hold: 
$X_{n} = \ln x_{n}$, $Y_{n} = \ln y_{n}$, 
$T = \ln \tau$, $A = \ln a$, and $B = \ln b$.
Here the phase $\Theta_{n}$ for $(X_{n}, Y_{n})$ is introduced as 
\begin{equation}
	\Theta_{n} = \arctan 
	\displaystyle \frac{Y_{n} - \ln \bar{y}}{X_{n} - \ln \bar{x}}.  
	\label{eqn:ud_theta}
\end{equation}
Based on eq.(\ref{eqn:Selkov_maxplus}), 
we obtain the bifurcation diagram of $\Theta_{n}$ 
as a function of $\tau$.
Actually, the red scatter plot in Fig.\ref{fig:BifDiagram_Selkov_ul_ov} 
shows the result.
As in the case of the negative feedback model, 
saddle-node bifurcation is retained 
in the max-plus system given by eq.(\ref{eqn:Selkov_maxplus}).
Here the saddle-node bifurcation point 
for eq.(\ref{eqn:Selkov_maxplus}), 
denoted by $\tau_{\text{m}}^{\ast}$, 
was numerically estimated 
as $\tau_{\text{m}}^{\ast} \approx 239.29993555\cdots$. 

\begin{figure}[t]%[h!]
	\centering
	\begin{minipage}{0.45\textwidth}%{0.23\textwidth}
			\includegraphics[width=\textwidth]{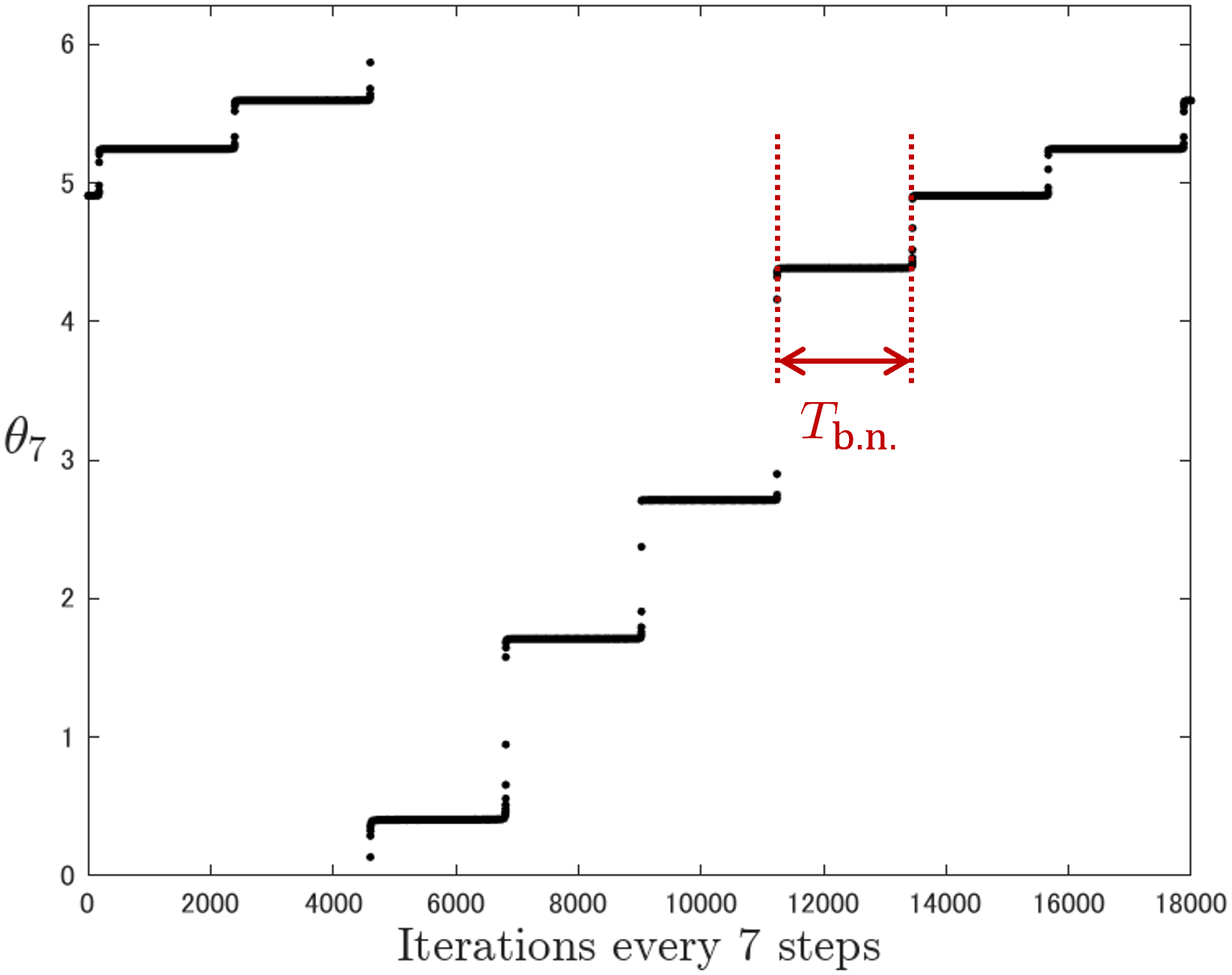}
			\centering
			(a)
	\end{minipage}
	%\hfill
	\begin{minipage}{0.45\textwidth}%{0.23\textwidth}
			\includegraphics[width=\textwidth]{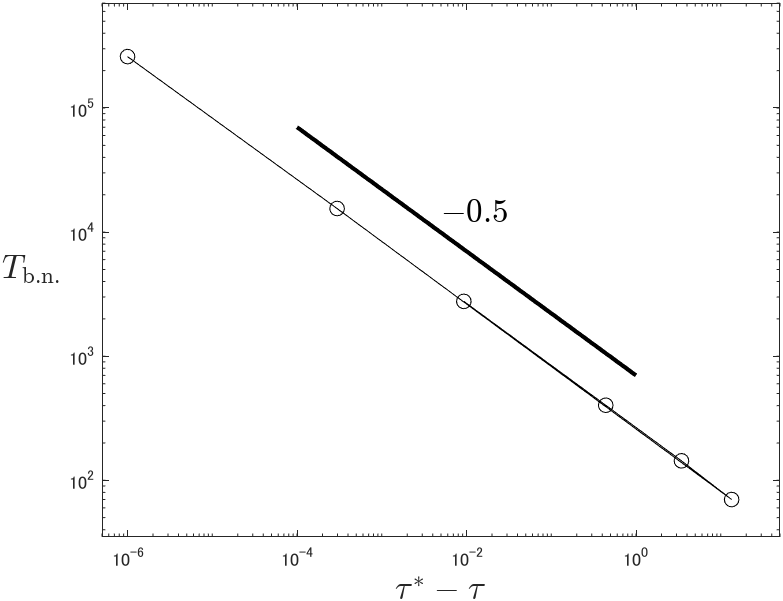}
			\centering
			(b)
	\end{minipage}
	\caption{\label{fig:ts_phase_Selkov_tau473} 
		(a) The bottleneck motion of $\theta_{7}$ 
		for $\tau \lesssim \tau^{\ast}$ obtained from 
		eq.(\ref{eqn:Selkov_tropical}).
		We set $\tau = 473.439$.
		(b) The scaling relation between 
		%the average time for passing bottlenecks, 
		$T_{\rm{b.n.}}$ and $\tau^{\ast}-\tau$.
		}
\end{figure}

\begin{figure}[ht!]%[h!]
	\centering
	\begin{minipage}{0.45\textwidth}%{0.23\textwidth}
			\includegraphics[width=\textwidth]{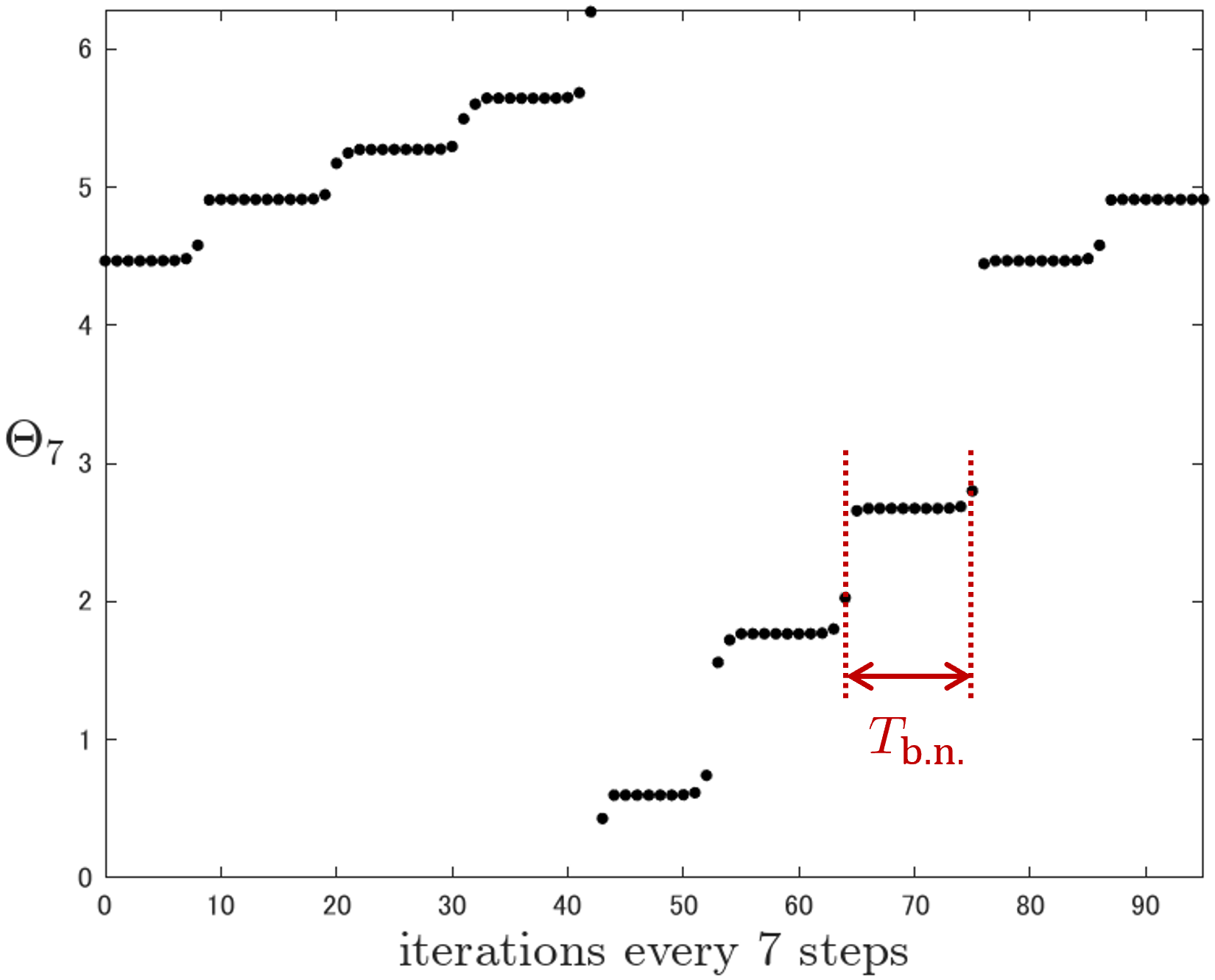}
					\centering
			(a)
	\end{minipage}
	%\hfill
	\begin{minipage}{0.45\textwidth}%{0.23\textwidth}
			\includegraphics[width=\textwidth]{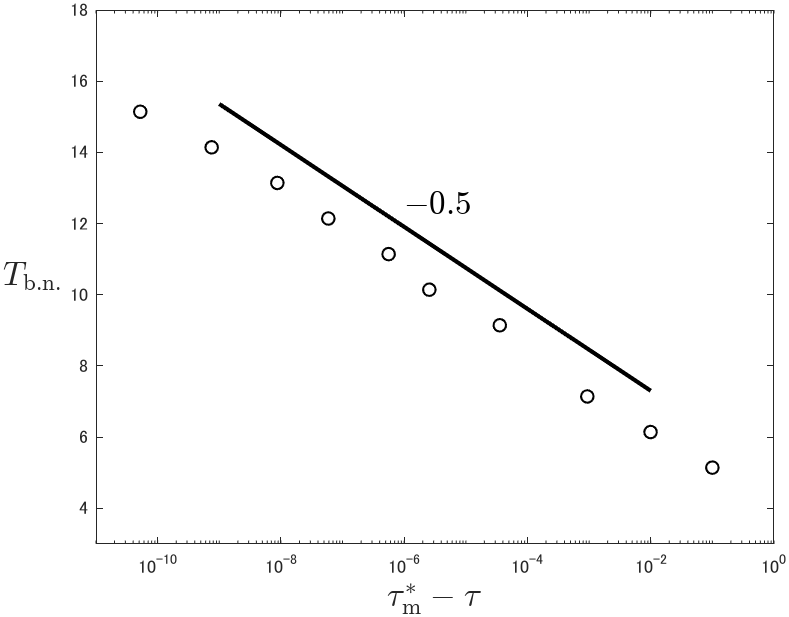}
			\centering
			(b)
	\end{minipage}
	\caption{\label{fig:ts_phase4_maxplus} 
		(a) The max-plus bottleneck motion of $\Theta_{7}$ 
		for $\tau \lesssim \tau^{\ast}$ obtained from 
		eq.(\ref{eqn:Selkov_maxplus}).
		We set $\tau = 239.299935$.
		(b) The logarithmic dependence of 
		%the average time for passing max-plus bottlenecks, 
		$T_{\rm{b.n.}}$ on $\tau_{\text{m}}^{\ast}-\tau$.}
\end{figure}

Figure \ref{fig:ts_phase_Selkov_tau473}(a) shows the phase drift and bottleneck motion 
for the phase every 7 steps, $\theta_{7}$, 
when $\tau \lesssim \tau^{\ast}$ in eq.(\ref{eqn:Selkov_tropical}).
As a scaling property for the average time 
to pass through the bottlenecks, denoted by $T_{\rm{b.n.}}$, 
we can confirm $T_{\rm{b.n.}} \sim (\tau^{\ast} - \tau)^{-0.5}$ 
from Fig.\ref{fig:ts_phase_Selkov_tau473}(b).
In the max-plus case of eq.(\ref{eqn:Selkov_maxplus}), 
the phase drift and bottleneck motion 
for the phase every 7 steps, $\Theta_{7}$, 
can be also observed as shown in Fig.\ref{fig:ts_phase4_maxplus}(a) 
when $\tau \lesssim \tau_{\text{m}}^{\ast}$.
The fact that the bottleneck motion is retained 
in the max-plus system is a common property 
in the Sel'kov model and the negative feedback model.
Moreover, as shown in Fig.\ref{fig:ts_phase4_maxplus}(b),  
we have newly found the logarithmic relation, 
$T_{\rm{b.n.}} \sim -\ln (\tau_{\text{m}}^{\ast} - \tau)$, 
in the max-plus case, 
which is different scaling relation from the discrete case. 

Comparing Fig.\ref{fig:ts_phase_Selkov_tau473}(b) 
and Fig.\ref{fig:ts_phase4_maxplus}(b), 
it is interesting that the ultradiscrete limit 
brings about the change of the scaling property for $T_{\rm{b.n.}}$.
The reason for the logaritmic dependence of $T_{\rm{b.n.}}$ 
on $(\tau_{\text{m}}^{\ast} - \tau) \equiv \lambda$ 
in the max-plus system 
can be explained as follows. 
The parameter $\lambda$ corresponds to the characteristic size 
of bottleneck threshold.
Important point is that the time evolution 
of the states becomes piecewise linear
by taking the ultradiscrete limit. 
Then the initial state approaches the bottleneck 
at the constant contraction ratio $r ( < 1 )$ 
for each time step.
Therefore, the time step $T_{\rm{b.n.}}$ satisfying 
$r^{T_{\rm{b.n.}}} \sim \lambda$ is what we desire.
Since $\ln r < 0$, we obtain 
$T_{\rm{b.n.}} \sim -\ln (\tau_{\text{m}}^{\ast} - \tau)$.
Note that we have also confirmed that this logarithmic dependence 
also holds for the max-plus negative feedback model.

Finally we comment on the possibility 
of general treatment for their dynamical properties 
by comparing the negative feedback model 
and the Sel'kov model.
For eq.(\ref{eqn:Selkov_maxplus}), 
considering a sufficiently large $T$ (or $T \to \infty$) 
and the following variable transformations, 
$\displaystyle X_{n}-\frac{A}{2} \to X_{n}$, 
$\displaystyle Y_{n}+\frac{A}{2} \to Y_{n}$, and 
$\displaystyle B - \frac{A}{2} \to B$,
we obtain the simplified max-plus Sel'kov model\cite{Ohmori2021,Yamazaki2021,Ohmori2022}, 
\begin{equation}
	\left\{
	  \begin{aligned}
  		X_{n+1} & = Y_{n} + \max(0, 2X_{n}), \\
	  	Y_{n+1} & = B - \max(0, 2X_{n}).
		\end{aligned}
	\right.
	\label{eqn:Selkov_simple_maxplus}
\end{equation}
On the other hand, 
regarding the negative feedback model\cite{Yamazaki2023,Ohmori2023c,Gibo2015}, 
\begin{equation}
	\left\{
		\begin{aligned}
			\frac{dx}{dt} & = y-x, \\
			\frac{dy}{dt} & = \frac{1}{1+x^{m}} - \frac{y}{b},
		\end{aligned}
	\right.
	\label{eqn:negative_continuous}
\end{equation}
from its tropically discretized one, 
\begin{equation}
	\left\{
		\begin{aligned}
			x_{n+1} & = \frac{x_n+\tau y_n}{1+\tau}, \\
			y_{n+1} & = \frac{y_n+\frac{\tau}{1+x_{n}^{m}}}{1+\frac{\tau}{b}}, 
		\end{aligned}
	\right.
	\label{eqn:negative_tropical}
\end{equation}
we can obtain the following simplified max-plus 
negative feedback model\cite{Yamazaki2023,Ohmori2023c}, 
\begin{equation}
	\left\{
	  \begin{aligned}
  		X_{n+1} & = Y_{n}, \\
	  	Y_{n+1} & = B - \max(0, 2X_{n}), 
		\end{aligned}
	\right.
	\label{eqn:negative_simple_maxplus}
\end{equation}
by the same treatment as the Sel'kov model.
Therefore, it is natural to adopt 
the following equation 
as a more general set of equations involving 
both eq.(\ref{eqn:Selkov_simple_maxplus}) 
and eq.(\ref{eqn:negative_simple_maxplus}), 
\begin{equation}
	\left\{
	  \begin{aligned}
  		X_{n+1} & = Y_{n} + \max(0, RX_{n}), \\
	  	Y_{n+1} & = B - \max(0, SX_{n}).
		\end{aligned}
	\right.
	\label{eqn:general_simple_maxplus}
\end{equation}
The dynamical properties of the max-plus system 
given by eq.(\ref{eqn:general_simple_maxplus}) 
will be reported elsewhere.

In conclusion, our research has demonstrated 
that the emergence of ultradiscrete states, 
induced by phase locking 
as a result of saddle-node bifurcation 
in discrete limit cycles, 
is not exclusive to the negative feedback model 
but is also observable in the Sel'kov model. 
Moreover, a comparison between the simplified 
max-plus models presented in eq.(\ref{eqn:Selkov_simple_maxplus}) 
and eq.(\ref{eqn:negative_simple_maxplus}) reveals 
a striking similarity, 
despite the apparent divergence in the corresponding 
original continuous models. 
Additionally, it has been found 
that the logarithmic relationship 
governing the average time required for a phase 
to pass through the bottleneck in the max-plus system 
stems from the piecewise linearization 
of the tropically discretized dynamical systems.
These findings lead us to propose 
that the occurrence of ultradiscrete states 
due to phase locking could be 
a prevalent characteristic in various other models.
%

% In conclusion, we have found that the emergence 
% of ultradiscrete states due to phase lock 
% caused by saddle-node bifurcation 
% in the discrete limit cycles is confirmed 
% not only in the negative feedback model 
% but also in the Sel'kov model.
% %
% Additionally, the simplified max-plus models 
% given as eq.(\ref{eqn:Selkov_simple_maxplus}) and 
% eq.(\ref{eqn:negative_simple_maxplus}) are quite similar 
% to each other, although the original continuous models 
% seem to be more different.
% %
% Furthermore, we have found that 
% the logarithmic dependence of the average time for the phase 
% to pass the bottleneck in the max-plus system 
% is caused by the piecewise linearization of the tropically discretized dynamical systems.
% %
% This conclusion suggests that the appearance 
% of ultradiscrete states due to phase lock 
% can be a common feature in some other cases.

\bigskip

\noindent
{\bf Acknowledgement}\\
The authors are grateful to 
Prof. M. Murata, Assoc. Prof. K. Matsuya, 
Prof. D. Takahashi, Prof. R. Willox, Prof. H. Ujino, 
Prof. T. Yamamoto, and Prof. Emeritus A. Kitada 
for useful comments and encouragements. 
This work was supported by JSPS
KAKENHI Grant Numbers 22K13963 and 22K03442.

\end{document}